# Machine Learning Optimization of E-Beam Transport for a Superradiant FEL

Amir Weinberg[1]*, Leon Feigin[2], Ariel Nause[2], Avraham Gover[1†]

[1]Center for Light-Matter Interaction, Tel Aviv University, Ramat Aviv, Israel

[2]Department of Physics, Ariel University

* amirwe@tauex.tau.ac.il

[†]gover@eng.tau.ac.il

30/7/26

**Abstract:** We present an optimization procedure using machine learning (ML) libraries for optimization of electron beam transport for maximal bunch compression and optimal operation of a bunched-beam Superradiant FEL. This is exemplified for the parameters of the 6MeV ORGAD Accelerator at Ariel University that is driving a THz Superradiant waveguide FEL. For superradiant emission (proportionally to the number of electrons squared), the bunch duration $\sigma_t$ at the undulator should be shorter than the optical period (2π/ω) of the radiation. Also, the beam transport optimization must confine the transverse dimensions of the beam to enter the undulator waveguide. The variables of the ML Bayesian optimization are the RF parameters and the currents of the coils and quads along the beamline. The beam dimensions and duration are provided from full 3D GPT simulations that are automatically driven by the ML exploration and exploitation algorithms. Twenty-five simulation iterations sufficed to arrive to an optimal beam transport design.

## I. Introduction:

We present an optimization procedure using machine learning libraries for electron beam transport through the entire beamline of the ORGAD RF Linac in the Schlesinger Center for Compact Accelerators in Ariel University [1,2], see Fig. 1. This compact accelerator (photo-cathode injector) serves a special kind of THz superradiant FEL. Contrary to conventional FEL, superradiant FEL is a source of coherent spontaneous emission of radiation – in the present context coherent synchrotron undulator radiation [12]. Such coherent intense radiation from an electron beam is possible if the beam is a short bunch, satisfying the superradiance condition:

$$f\sigma_t \ll 1 \qquad (1)$$

where f is the frequency of the emitted radiation and $\sigma_t$ is the standard deviation of the beam temporal distribution. Under this condition all electrons of the bunch emit in phase radiation wavepackets that interfere constructively to produce a radiation pulse of energy proportional to $N^2$, where N is the number of particles in the bunch (while spontaneous emission energy is proportional to N). This bunched beam superradiance is analogous to Dicke's superradiance emission from excited atoms [13].

The compact moderate energy (6 MeV) accelerator of the Superradiant FEL shown in Fig. 1, is a hybrid S-band (2856 MHz) photo injector [3] composed of 3.5 standing-wave gun cells integrally connected to subsequent 9 traveling-wave gun cells, hence, the name

“hybrid”. In the first 20cm standing-wave acceleration section the particles can reach energy of 6.5Mev. The subsequent traveling wave section, 40cm long, modulates the beam to acquire negative energy chirp. Following the gun, the beam freely drifts through a beam line up to the entrance to the undulator 4 m away from the cathode, where due to the free space energy dispersion, it is intended to compress into a short bunch at the center of the undulator. The undulator at the end of the beamline is a planar Halbach undulator - 80cm long, enclosing a waveguide section of cross-section 13X6.5 $mm^2$ which guides the THZ Superradiant radiation generated by the undulator.

The design optimization of the electron beam transport through the beamline requires adjustment of the magnetic electron-optical elements – coils, quadruples along the wiggler and the accelerator RF-field parameters in order to achieve minimum bunch duration in the undulator while keeping the focused beam transverse dimensions small enough to pass through the entire beamline and extrude the aperture of the waveguide at the entrance of the undulator. In this work, the simulations of the electron beam trajectories were carried out using the General Particle Tracer code (GPT) [4]. GPT requires 3-D maps of all magnetic field along the beamline in order to propagate a beam of given quality parameters (energy spread, emittance) generated at the cathode. Some of these fields are generated by model computation of GPT (undulator, coils, quads), others are supplied by the user, based on measurement characterization of the gun elements (RF field and magnetic field distributions in the gun). The optimization process consists of adjusting more than ten parameters of electron-optical element currents and gun RF intensity and phase in order to attain a target of minimal electron beam bunch duration (1). For a Gaussian beam distribution, the SR emission is proportional to the bunching parameter $|M_b|^2$ [12] which is the figure of merit for maximal SR energy emission:

$$|M_b|^2 = e^{-(2\pi f \sigma_t)^2} \qquad (2)$$

In the superradiance condition, $f\sigma_t \ll 1$, this parameter reaches a value of unity. In the opposite case the Superradiant emission vanishes.

Manual optimization of the multiple variable parameters to achieve the optimal goal target along with the limitations of the beam transport through apertures, is an elaborate process. Here we present an optimization method using ML (machine learning) libraries, Bayesian optimization [9,10,11]. The presented optimization shows significant improvements as well as improvements in time consuming and parameter browsing relative to previous optimizations methods [6] without the ML.

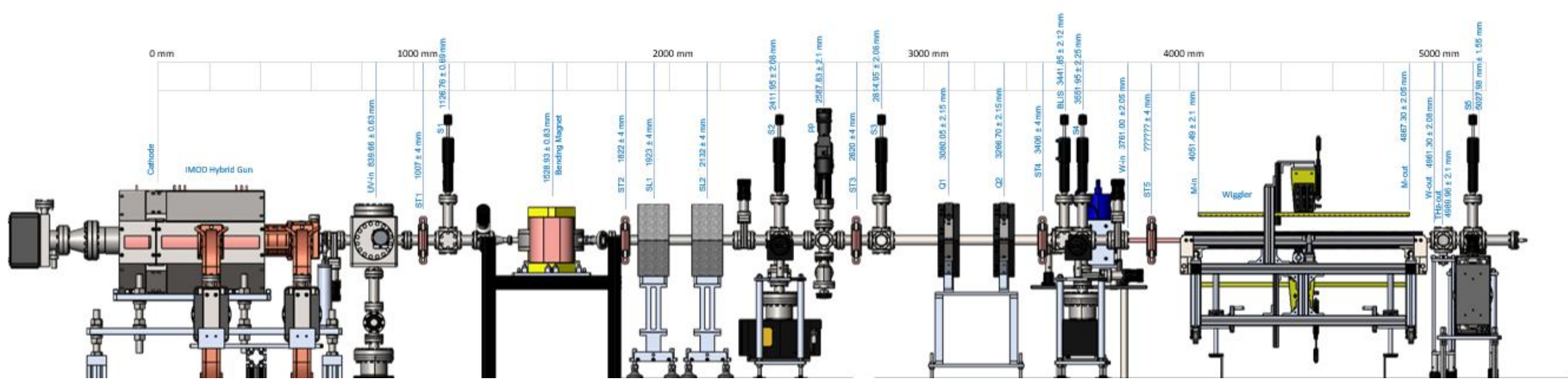

*Figure 1. The ORGAD accelerator and THz SR FEL at Ariel University, comprised of a photocathode gun injector, two focusing solenoids (gray), two quadrupoles (black)black, and an undulator section starting at 4.051m and ending at 4.8m from gun cathode.*

## II. OPTIMIZATION METHOD

Full 3D simulations of the hybrid RF-gun, including space-charge effects, were performed using GPT. The simulations start from the cathode, with randomized normal distribution of particles using a charge of 50pC and beam quality parameters listed in Table 1. In the simulations, the accelerator gun was modeled using a realistic 3D field map of the gun RF cavities imported from HFSS computations. The magnetic fields of the gun coils, as well as the beamline lattice (focusing solenoids and quadruples), were measured experimentally and imported to the beamline simulation. Public domain software programs package - LUME-FEL [7] including object libraries LUME-GPT and XOPT [8], were integrated with the simulation. LUME-GPT is a python wrapper for GPT, and the XOPT library provides the Bayesian optimization algorithm, which is suitable for exploration of the parameter space.

Table 1: Electron beam parameters

| | Value |
|---|---|
| Charge Q | 50 pC |
| Initial Particle distribution $\sigma_t$ | 1.5pS |
| Initial transverse distribution $\sigma_{x,y}$ | 0.1mm |
| Number of macro particles | 2000 |
| Acceleration energy | 6 MeV |

The entire simulation was converted to a function with inputs and outputs to run in the python language. The function inputs are the gun parameters and beam-line lattice currents. The function outputs are the e-beam duration $\sigma_t$ in the middle of the undulator (4.4m from the cathode) and the e-beam transverse dimensions $\sigma_{x,y}$ (4m from the cathode) at the entrance to the undulator.

The optimization procedure is managed using a script of VOCS (variables, objectives and constraints) [8]. In our example of the beamline transport optimization, the script variables are the currents of two solenoids and two quadruples, each is given a range of current values to scan. The objective in our example is to minimize the duration of the beam bunch $\sigma_t$ and constrain the transverse dimensions of the beam $\sigma_x < 3mm$, and $\sigma_y < 1.5\ mm$ at the entrance to the undulator waveguide.

The Expected Improvement Generator [8] algorithm that we used from the library XOPT, is a Bayesian optimization method that solves single objective problems. The algorithm can balance between exploration and exploitation. Exploration means that the ML algorithm looks for new parameters in new locations of the parameter space. Exploitation means that the Bayesian algorithm scans for new parameter values in explored locations of the parameters space, trying to refine the results.

Based on the drawing of Fig. 1 we define the variable parameter space as follows. There are four variable parameters in the gun section and their values range are listed and given in Table 2.

- RF Electric field phase at the timing of the photocathode injection, in radian.
- RF Electric field relative intensity - MF (multiplication factor relative to a nominal reference).
- Currents of gun coils embedded in the iron wrapping of the gun $I_C$, $I_F$

*Table 2. Variable parameters of the gun section and their values range*

| | From | To |
|---|---|---|
| RF Phase at cathode $\Phi$ | 0.6[rad] | 1[rad] |
| MF of E-field on cathode | 0.6 | 1 |
| center gun coil current – $I_C$ | 60[A] | 120[A] |
| Front gun coil current – $I_F$ | 60[A] | 120[A] |

There are four variable parameters in the beamline section and their values range are listed in Table 3

*Table 3. Variable parameters of the beam-line lattice and their values range.*

| | From | To |
|---|---|---|
| First solenoid current $I_{SL1}$ | 0.5[A] | 4[A] |
| Second solenoid current $I_{SL1}$ | -4[A] | -0.5[A] |
| First Quadrupole current $I_{Q1}$ | -4[A] | -0.5[A] |
| Second Quadrupole current $I_{Q2}$ | 0.5[A] | 4[A] |

The optimization process was carried out in two steps. First, we optimized the gun parameters for the target goal (minimal $\sigma_t$), nulling the parameters of the beamline. Then we used the selected gun parameters as a starting point for optimization of the beamline parameters.

The entire optimization process consisted of only twenty-five runs of the beamline simulation. In the first ten runs of the simulation the algorithm generated random input parameters and based on the outputs of the simulation created a map of parameters space. In the next fifteen simulations the Bayesian algorithm, based on the map created, tries to achieve it's objective.

## III. Parametrization of the magnetic and RF field distributions

While GPT simulation of electron trajectories requires full 3-D maps of the magnetic field distribution in the gun, in the transport beamline and in the undulator and a 3-D map of the RF fields in the gun cavities, the ML algorithm requires specification of a finite number of variable discrete parameters. We show how it is done for the specific example of ORGAD accelerator.

The hybrid photocathode gun is constructed in such a way that a standing wave RF field is excited in the first 3.5 cells of the gun, providing acceleration of the beam from the cathode up to the end of this section and the next 9 traveling wave RF gun cells provide energy chirp modulation to the beam. The integral 12.5 cavities structure is fed through RF waveguides connected right at the connection between the two sections (see Fig. 1) so that the RF field phase between the two sections is set fixed to be 90° [1]. HFSS code [5] simulation of the RF field distribution was performed for the given geometrical structures of the manufactured gun. This distribution is fed into GPT input file as a reference field map calculated for a given RF power from the klystron. The GPT simulation program requires two inputs to determine the actual 3-D field distribution map: a multiplication factor MF relative to the reference input RF power and the phase Φ of the RF field relative to the timing of the laser incidence on the cathode.

The magnetic field distributions of the two focusing coils SL1, SL2 of the beamline were provided by the manufacturer (Danfysik) and fed as a 3-D maps into GPT. These coils have no iron core and their magnetic field distribution intensity is therefore proportional to the coil current and uniquely determined by the coil current parameter.

The two quads on the beamline are home manufactured. Their gradient intensity and dependence on their excitation current were determined from lab measurements and were fed into GPT input files, characterized by a single parameter of excitation current.

The magnetic field distribution in the undulator is a standard function of GPT given the dimensions, period and magnetic strength of the undulator magnets: $L_w = 0.8\,m, \lambda_w = 2\,cm, B_w = 0.49\,Tesla$ . It is kept constant for all runs and was not a variable of the optimization.

The parametrization of the magnetic field distribution of the gun coils is more intricate. These coils are embedded inside an iron enclosure, wrapping the gun, which is shown in Fig. 1 as a gray area. The location and geometry of the coils in the embedded gun are shown in Fig. 2. The strong axial magnetic field of the gun coils keeps the beam focused along the gun section, keeping it from expansion due to the strong transverse space-charge field. The internal structure of the gun coils ensemble is composed of three coil units: Bucking coil (B), Center coil (C) and Front coil (F) located respectively at (-83mm, 81.5mm, 341mm) relative to the cathode position. They are composed respectively of two, three and three identical solenoids. The solenoids in each coil unit are connected in series, so that the field distribution along the gun is determined by three current parameters $I_B, I_C, I_F$. The magnetic field distribution of a single solenoid was measured in the laboratory and was used to calibrate the GPT model of a single coil unit. Given the location of the three coils, GPT constructs a 3-D map of the magnetic field along the entire gun as a superposition of the solenoid fields. The magnetic field along the gun axis was measured earlier (red line in Fig.

2). For the calibration of the GPT model coils, a particular choice of reference coil currents $I_B = -178.11, I_C = 107.6, I_F = 98A$ (see Fig. 2) reproduces a field distribution (dashed blue line) that replicates the measured distribution (red line).

The bucking coil behind the cathode is intended to cancel the field produced by the other coils and null the field on the cathode in order to avoid emittance growth effect of the beam upon emission. This must be kept also when the coil currents are changed in the optimization process. A special algorithm is included in GPT that adjusts $I_B$ to null the field on the cathode for any choice of $I_C, I_F$. Therefore, GPT can construct the magnetic field distribution based on only two current parameters $I_C, I_F$ that are used as variables in the transport optimization process.

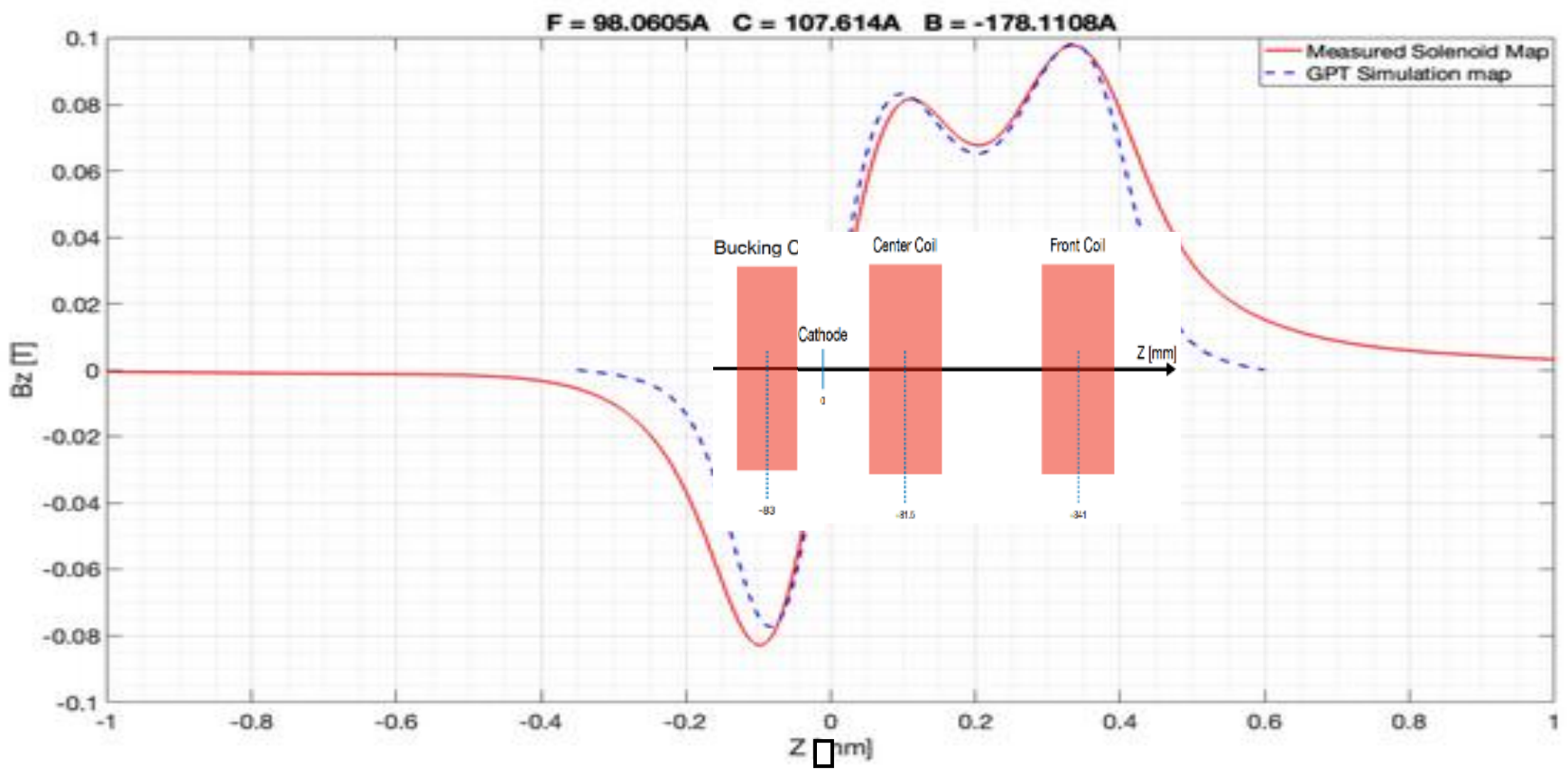

*Figure 2: Axial magnetic field distribution along the gun for a specific set of internal coils currents. Three coils are located at locations (with respect to the cathode): bucking coil -83mm, center coil 81.5mm and front coil 341mm from the cathode. The field on the cathode is set null to avoid emittance growth.*

## IV. ML Optimization of the ORGAD beamline for enhancement of Superradiant emission

We demonstrate the application of the ML procedure for optimization of the THz Superradiant FEL based on the ORGAD hybrid gun accelerator [1]. The beam transport simulations were optimized with the ML procedure in two steps. In the first optimization step the ML algorithm scanned only over the four parameters of the gun section (Table 2) keeping the parameters of the beam line (Table 3) constant (null currents). The GPT simulations were carried out with the initial beam parameters given in Table 1. The limitations set on the beam transverse dimensions were ignored and the optimization target parameter - the bunch duration at the center of the undulator $\sigma_t$- was derived out of the GPT simulation at each of 25 ML optimization steps

The optimization algorithm scanned over the parameter values ranges of Table 2 and sampled 25 points in the four-dimensional space of the gun parameters (Φ, MF, $I_C$, $I_B$). The ML exploration and exploitation process starts with 10 random selections of parameter values, which get improved with the exploitation steps, however, the set of parameter values of the last step is not necessarily the best, and the user has freedom to examine and choose another set from the provided sampling results. For better exploitation of the optimization results, the ML algorithm of LUME-FEL creates a continuous probability function of the target $\sigma_t$ in the four-dimensional space based on the sampling points and their prediction uncertainty [9]. This function is visualized in a color map by presenting two sub-space cuts: (a) phase-intensity (Φ-ML) and (b) gun coils ($I_C - I_B$). Fig. 4 shows these maps for the set of parameter values that corresponds to the last of the 25 iteration steps. The projection of all 4-D sampling points on this cut is also shown as yellow dots in the two maps. It is possible to view the projection of these points in any other cuts of the 4D space. Minding the color maps and checking the beam trajectories in the corresponding GPT simulation, a set of four optimal gun parameters were selected, marked with a "star" and listed in Table 4a.

The second optimization step was applied on the beamline lattice parameters (Table 3): the two solenoids' currents ($I_{SL1}, \ I_{SL2}$) and the quadrupole doublet ($I_{Q1}, \ I_{Q2}$). The gun section parameters were kept constant in the GPT simulations. These are the four selected parameters of the first step listed in Table. 4a. The results of the second ML optimization process are displayed in Fig. 5 in two sub-space cuts of the four-dimensional parameters space: (a) ($I_{SL1}, \ I_{SL2}$) and (b) ($I_{Q1}, \ I_{Q2}$). In this run we mind the constraints on the transverse dimensions of the beam $\sigma_x$ and $\sigma_y$; the infeasible samples that do not satisfy the transverse dimensions constraints are marked as empty circles and the feasible results are solid circles.

The color code maps of the optimized parameter space in Fig 5 were constructed based on the feasible solutions (solid circles). Out of these solutions, the selected best set of parameters is listed in Table 4b. Minding this result and the color map as a starting point, we performed independent GPT simulations of the entire setup for the beam parameters of Table 1 for beam charge of 50pC and beam energy of 6 MeV. The independent GPT simulations led to selection of a better set of beamline parameters listed in Table 4b in parentheses and marked in Fig. 5 with a "star". The results of the GPT simulation for this set of parameters are shown in section V (Figs. 6-8) and suggest achievements of compression of the electron beam bunch to optimally short duration of $\sigma_t$=146fs at the middle of the undulator (see top panel of Fig. 8). This corresponds to a satisfactory value of the bunching parameter (Eq. 2) $|M_b|^2 = 0.81$ for radiation at frequency of 0.5THz. The expected transverse dimensions $\sigma_x$=4mm and $\sigma_y$=0.4mm at the entrance to the undulator-waveguide section (second and bottom panels of Fig. 8) satisfy the waveguide aperture constraints.

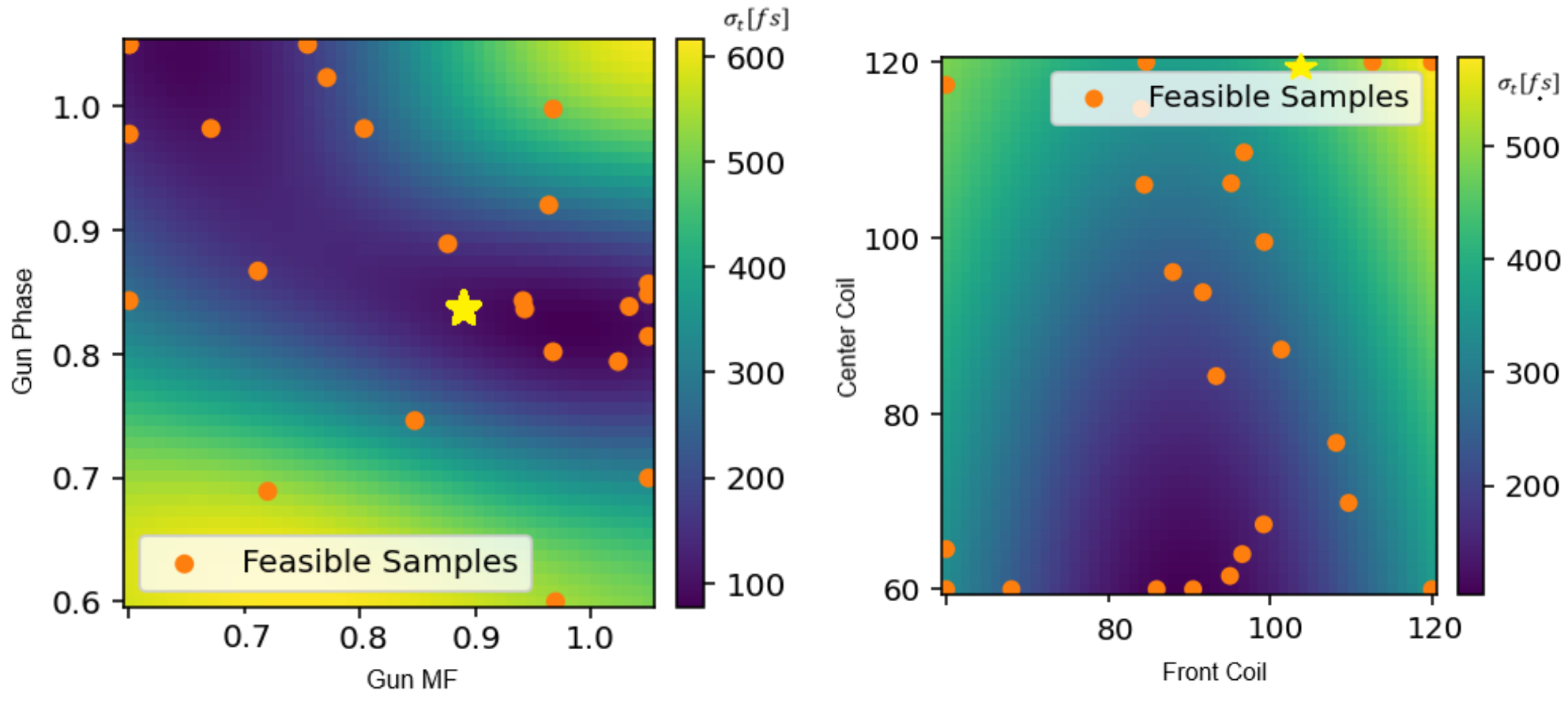

*Figure 4: Color code maps of the first optimization step over the gun parameters. The figure displays the 4D function of $\sigma_t$ for the four gun parameter values of the last ML iteration. It is displayed in terms of two 2D cuts of the 4D parameters space: (a) Gun RF phase and field intensity Multiplication factor ($\Phi$ -MF), (b) Gun front and center coils ($I_C - I_B$). The projection of all 25 sampling points on the chosen cut plane is shown as yellow spots. Marked in yellow stars - the chosen parameters.*

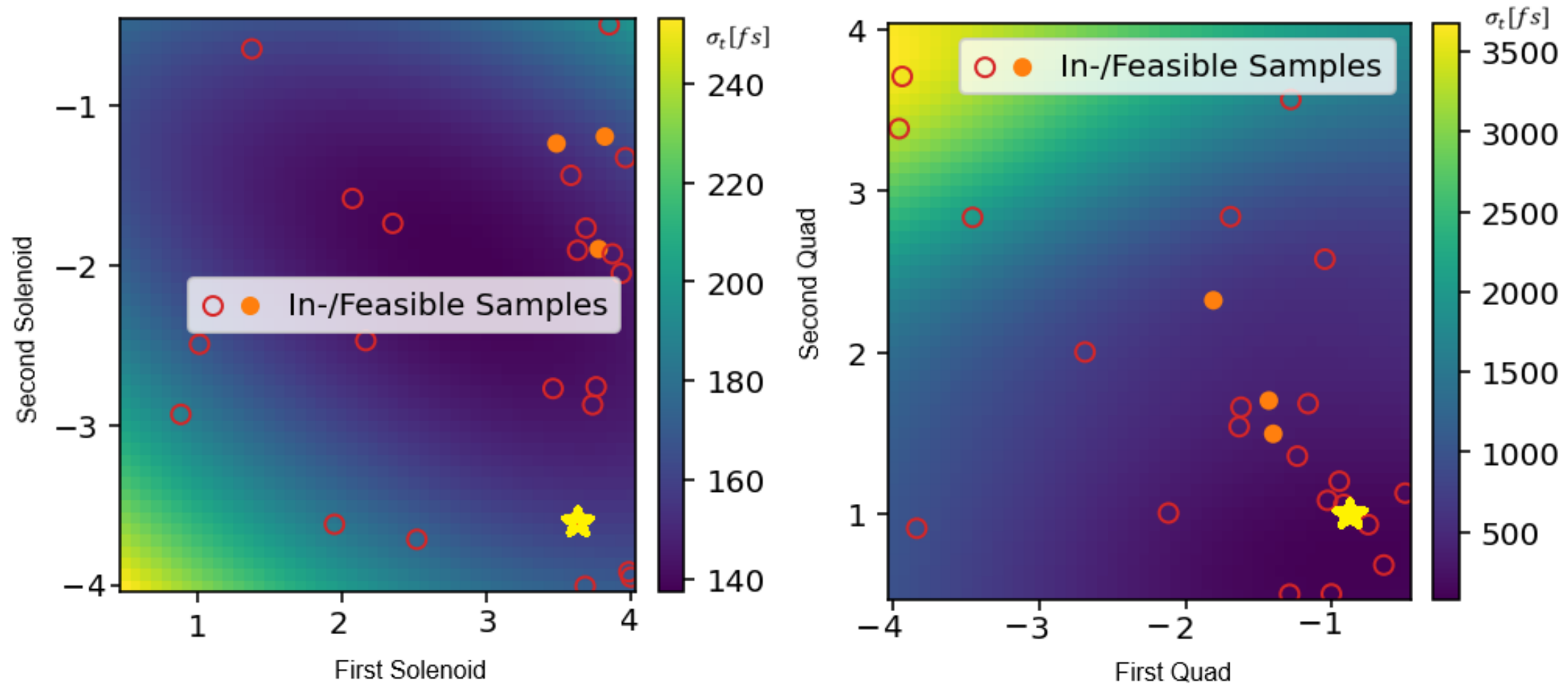

*Figure 5: Maps of the results of the second optimization procedure over the beamline lattice, (a) the solenoids ($I_{SL1} - I_{SL2}$), and (b) the quads ($I_{Q1}$-$I_{Q2}$). The open circles correspond to non-feasible sampling points that do not satisfy the transverse dimension constraints.*

*Table 4: Gun and beamline parameters selected in the ML optimization process*

| a. gun parameters | | b. Beamline parameters | |
|---|---|---|---|
| Phase Φ=47.56deg | MF=0.89 | $I_{SL1} = 3.6A\ (4A)$ | $I_{SL2} - 3.6A\ (-4A)$ |
| Center $I_C = 119A$ | Front $I_F = 103.6A$ | $I_{Q1} = -0.6A$ | $I_{Q2} = 1A\ (0.2A)$ |

V. GPT numerical computation and confirmation

We display the beam simulation results for the selected set of parameters (Table 4 with the parameter values in parentheses) in Figures 5 to Figure 8. Figure 5 shows the evolution of the kinetic energy of the beam in the gun section. The selected phase Φ does not correspond to ideal on-crest acceleration followed by ideal energy chirp modulation. Because in the hybrid gun, the phase between the acceleration and the modulation sections is fixed and is not a degree of freedom, the values of the degrees of freedom *Φ -MF* represent a compromise of acceleration and modulation (chirping). Fig. 5 shows that the compromised phase choice Φ=47.56deg leads to beam acceleration to 6.5 MeV in the first twenty centimeters of the 3.5 cell standing-wave gun section and slight deceleration in the modulation section, settling at energy of 6 MeV at the end of the gun section at z=0.6m. Yet, good negative energy chirp is produced for the chosen phase in the modulation section of the gun as depicted in Fig. 7, showing the energy-time phase-space of the beam at the end of the gun section. The displayed negative energy chirp generated in the TW (traveling wave) section of the RF-gun is responsible for creating a waist at the desired location downstream in the undulator, as shown in the top panel of Fig. 8.

In Figure 8 we display the evolution of the electron beam bunch dimensions along the entire beamline. The top panel displays the e-beam duration $\sigma_t$ , starting at value of $\sigma_{t0} = 1ps$ and arriving at a minimal value of $\sigma_t$=146fs at the middle of the undulator. The second panel displays the evolution of the transverse dimensions of the e-bea $\sigma_x$, reaching the values of $\sigma_x$=4mm at the entrance to the waveguide in satisfaction of the ML optimization constraint. Correspondingly, also in the y dimension, the transport of the beam satisfies well the ML optimization constraint as depicted in the plot of $\sigma_y$ in the third panel of figure 8. Here the focusing property of the undulator keeps the beam well confined within the waveguide. Figure 9 shows the particle trajectories for the parameters which were chosen from the optimization, small percentage of the particles is found outside the red markings of the waveguide, which may be tolerable.

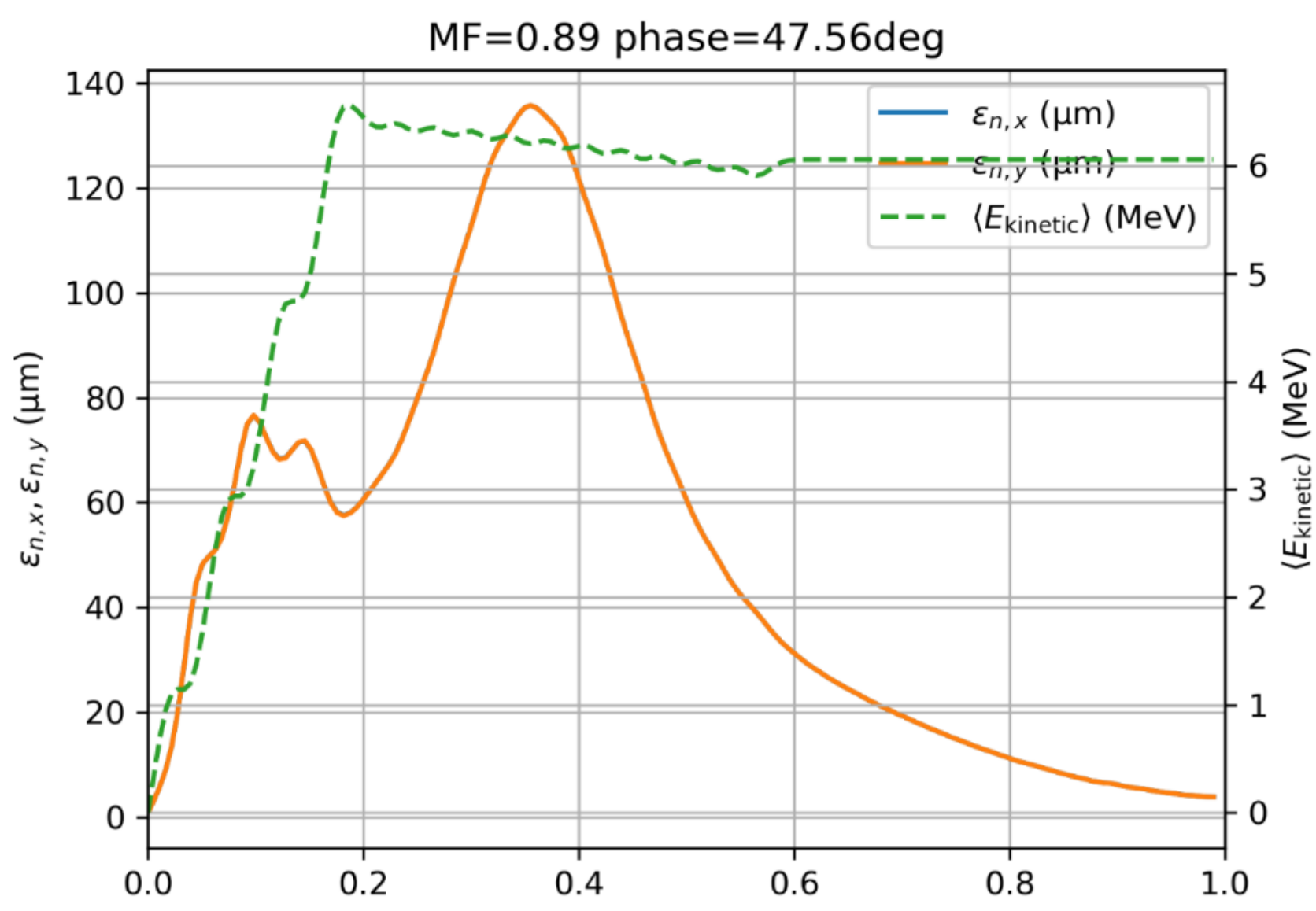

*Figure 6: Plot of the beam kinetic energy (in green) and emittance (in orange) along the gun section*

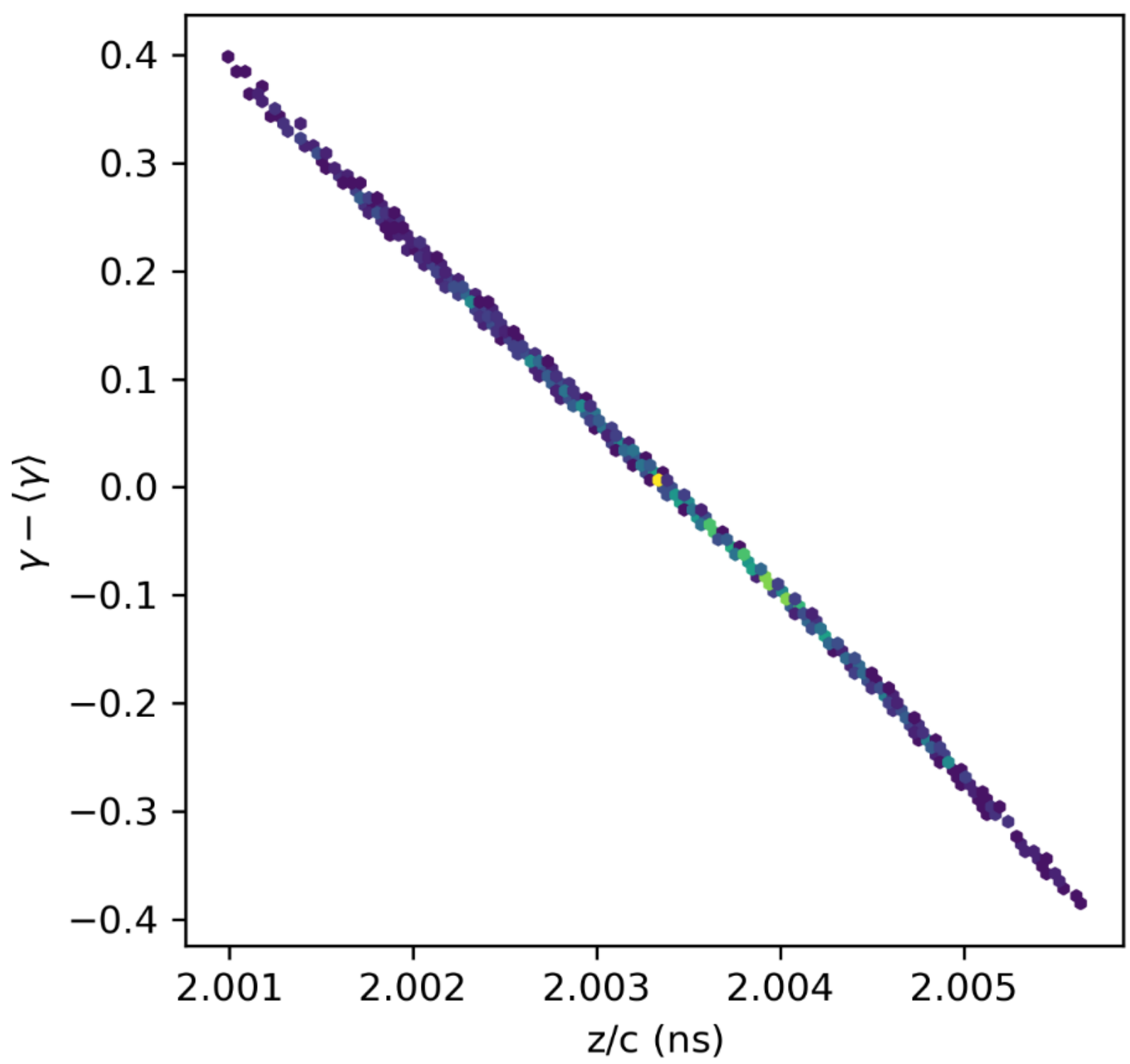

*Figure 7: Longitudinal phase space of the beam at the end of the accelerator gun section, 0.6m from cathode. The incremental energy along the beam bunch shows a negative chirp.*

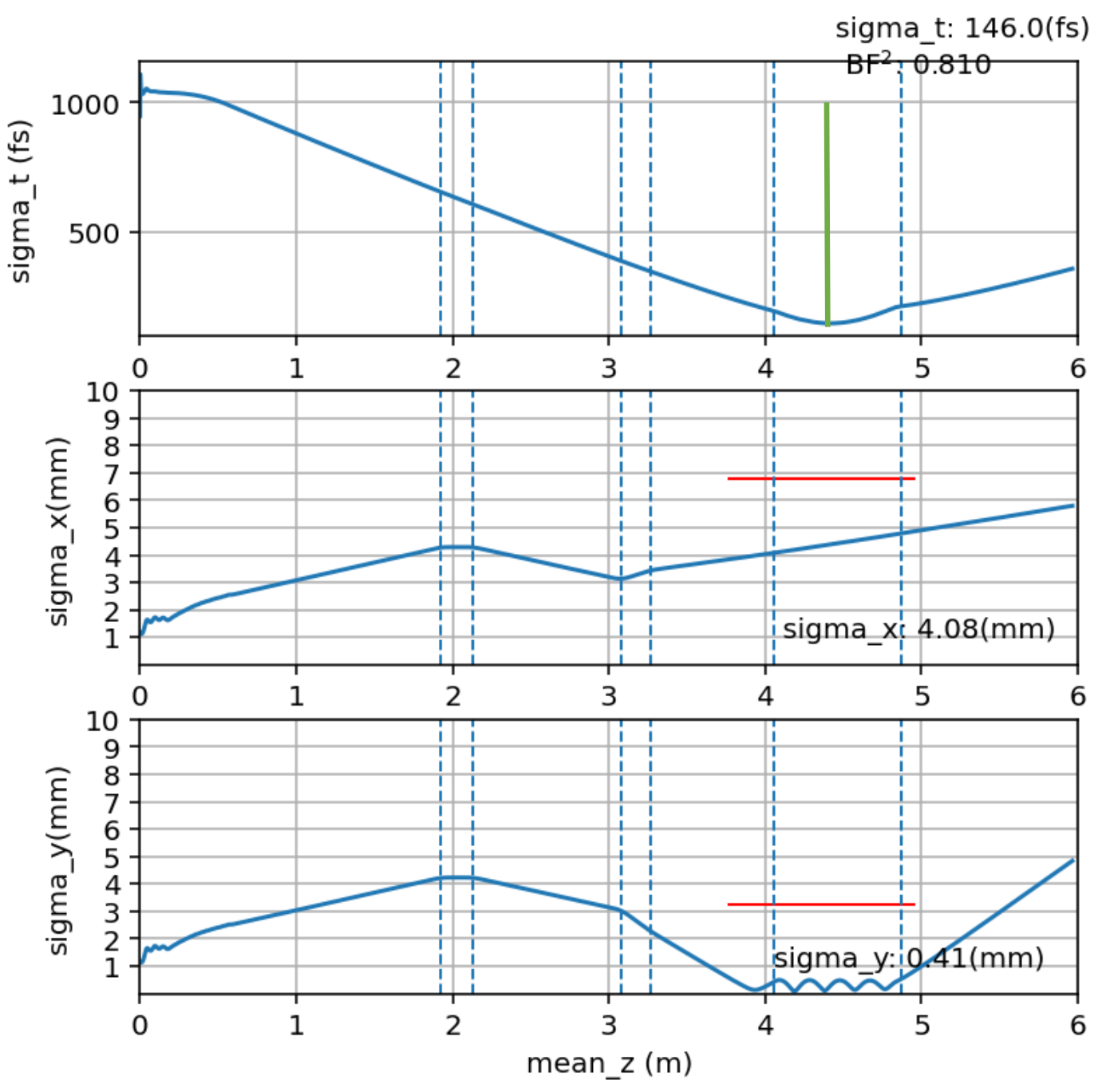

*Figure 8: Plots of $\sigma_t$, $\sigma_x$, $\sigma_y$.The bunch duration is $\sigma_t$=146.1fs in the middle of the undulator, this corresponds to bunching factor$|M_b|^2 = 0.81$ (for radiation frequency of 0.5THz). The*

*blue dashed lines mark the positions of the focusing solenoids, the quad doublet and the undulator section. The red line marks the waveguide borders.*

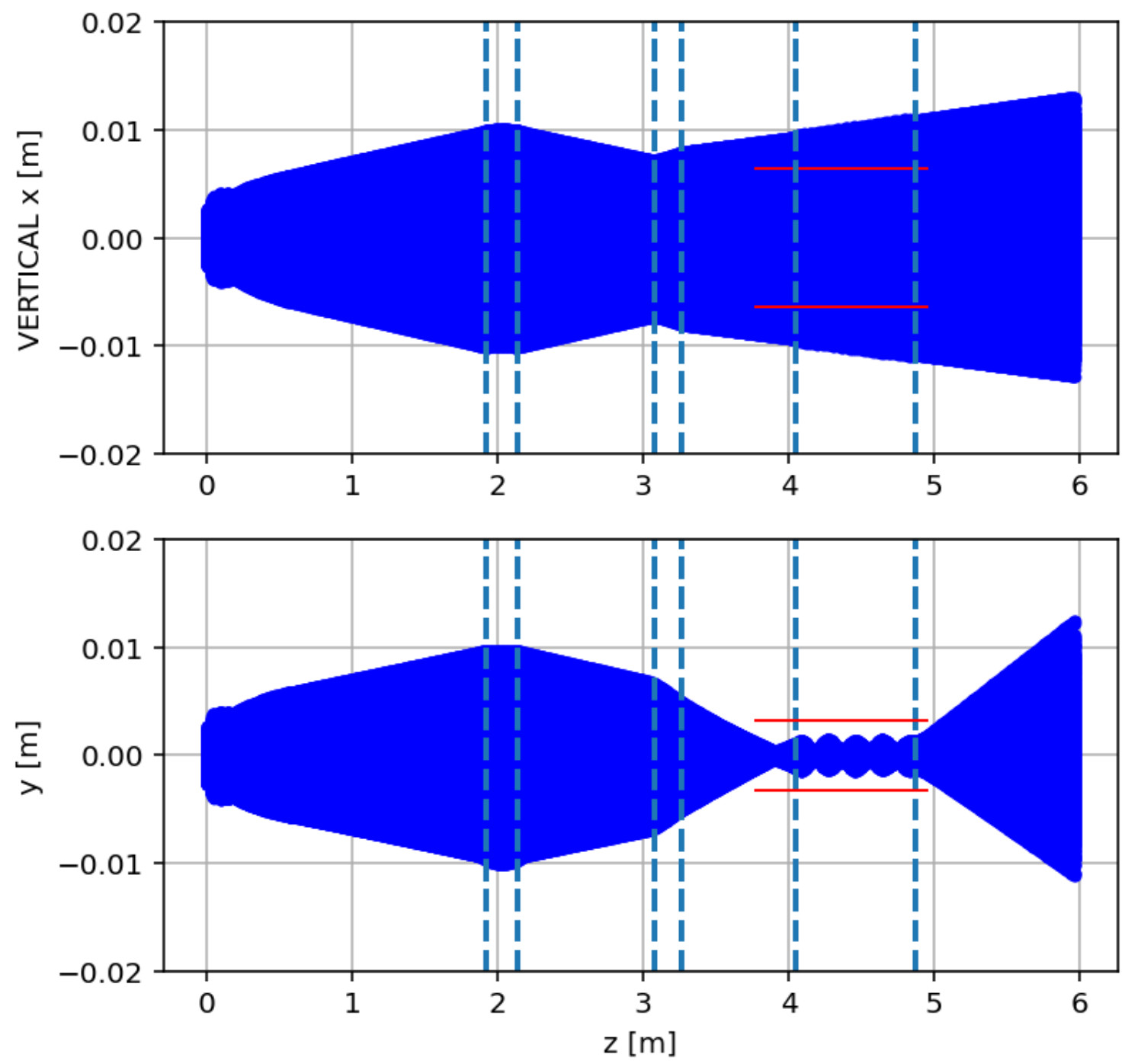

*Figure 9: Plot of the particle trajectories. The first four dashed lines mark the location of the focusing elements. Starting at the cathode, the beam diverges, it starts converging at 2m from the cathode due to the solenoids, and at 3m it converges (x) and diverges (y) by the quad doublet. The red lines mark the waveguide borders.*

The overall optimization time of the presented example on a laptop computer was less than ten minutes. This is quite a short time for running optimization, especially in comparison to our previous human directed optimizations of the ORGAD beamline.

## V Conclusions

The ML optimization procedure presented here, based on realistic consideration of measured field maps of RF fields, gun coils and beamline lattice optics, attains optimal results in short time for the parameters of Ariel's THz-FEL. This demonstrates the efficiency of this ML tool in designing and optimizing compact accelerators and FEL devices in general. The Machine learning libraries in the optimization method used in this work run the GPT beamline simulations only over twenty-five iterations. There was no need to build a large dataset as is usually done with ML. For the presented example of ORGAD RF-LINAC with beam charge of 50pC, the optimization target of minimal bunch duration at the center of the undulator attained is $\sigma_t$=146.1fs, corresponding to a bunching factor value $|M_b|^2 = 0.81$ for radiation frequency of 0.5THz. This is good enough for using the derived variable setup parameters as a starting point in a practical refined design of a Superradiant FEL at this frequency range.

**VI Acknowledgement**
The author(s) declare financial support was received for the research, authorship, and/or publication of his article. We acknowledge support by the Israel Science Foundation through grant 1705/22.